\documentclass[prl,twocolumn]{revtex4}
\usepackage{amsfonts}
\usepackage{mathrsfs}
\usepackage{amsmath}
\usepackage{amssymb}
\usepackage[medium]{titlesec}
\usepackage{bm}
\usepackage[normalem]{ulem}
\usepackage{extarrows}
\usepackage{slashed}
\usepackage{isodateo}
\usepackage{graphicx} 
\usepackage{xcolor}
\usepackage[utf8]{inputenc}
\usepackage[bookmarksnumbered=true,bookmarksopen=true]{hyperref}
 \hypersetup{colorlinks,%
             linkcolor=[rgb]{0,0.3,0.6}, %
             citecolor=[rgb]{0,0.3,0.6}, %
             urlcolor=[rgb]{0,0.3,0.6}}
\usepackage{indentfirst}
\renewcommand{\FR}[2]{\displaystyle\frac{\,{#1}\,}{#2}}

\newcommand{\n}{\nonumber}

\graphicspath{{fig/}}

\def\bge{\begin{equation}}
\def\ede{\end{equation}}
\def\bga{\begin{aligned}}
\def\eda{\end{aligned}}
\def\bgp{\begin{pmatrix}}
\def\edp{\end{pmatrix}}
\def\bgs{\begin{subequations}}
\def\eds{\end{subequations}}
\newcommand{\order}[1]{\mathcal{O}({#1})}
\def\di{{\mathrm{d}}}

\def\pd{\partial}

\def\la{\langle}\def\ra{\rangle}

\def\ep{\epsilon}

\newcommand{\wh}[1]{\mkern 2mu \widehat{\mkern-2mu#1\mkern-2mu}\mkern 2mu}
 
\newcommand*{\vcenteredhbox}[1]{\begingroup
\setbox0=\hbox{#1}\parbox{\wd0}{\box0}\endgroup}

\begin{document} 

\title{Observing Eccentricity Oscillations of Binary Black Holes in LISA}

\author{Lisa Randall}
\affiliation{Department of Physics, Harvard University, 17 Oxford St., Cambridge, MA 02138, USA} 

\author{Zhong-Zhi Xianyu}
\affiliation{Department of Physics, Harvard University, 17 Oxford St., Cambridge, MA 02138, USA}  
 
\begin{abstract}
The tidal force from a third body near a binary system could introduce long-term oscillations in the binary's eccentricity, known as Kozai-Lidov oscillations. We show that the Kozai-Lidov oscillations of stellar-mass black hole binaries have the potential to be observed by LISA. Detections of such binaries will give insights into  binary formation channels and also provide an important benchmark of observing Kozai-Lidov oscillations directly.
 
\end{abstract}

\maketitle

LIGO detections of gravitational waves (GWs) have revealed the existence of a potentially large number of stellar-mass binary black holes (BBHs) \cite{LIGOScientific:2018mvr,LIGOScientific:2018jsj}.  However, their origin, or formation channels, remains a crucial open question. Current theories suggest that such BBHs can form either in  isolation  or through  so-called dynamical formation channels that might occur in dense regions. 

One way to differentiate the two channels is to measure the statistical distributions of binary parameters, including the spin alignment and the eccentricity. It has been shown that LIGO and LISA has the potential to resolve many stellar-mass BBHs and to measure their spin alignment as well as orbital eccentricity \cite{Nishizawa:2016jji,Nishizawa:2016eza,Breivik:2016ddj,Rodriguez:2016vmx,Randall:2017jop,Randall:2018nud}. 

However the detailed distribution of parameters like the eccentricity may depend on many environmental factors that could introduce large uncertainties in making theoretical predictions. In addition, the distribution of the eccentricity produced in different dynamical channels may be degenerate making new classes of observations that shed light on formation channels quite valuable.

In \cite{Randall:2018lnh} we introduced the idea of measuring directly the barycenter motion of the BBHs due to the gravity of a nearby third body, and a measurement of the large orbital motion of the binary may serve as a direct measurement of the nearby mass density,  thus providing direct information about the BBH environment. See also \cite{Meiron:2016ipr,Inayoshi:2017hgw,Robson:2018svj} for similar works. In this letter, we introduce another distinct source of BBHs in the dynamical formation channel where the influence of the third-body is manifested as the explicit eccentricity oscillation in LISA. The Kozai-Lidov (KL) mechanism \cite{Kozai:1962zz,Lidov:1976qhg} explains how the tidal force from a nearby third body has the effect on the binary that the the binary orbit's inclination and the eccentricity undergo long-term exchange. We will show that in some cases, this time dependence of the eccentricity from KL oscillations could be observed in LISA. In particular, an increasing of eccentricity in time would be a smoking-gun signature of BBHs undergoing KL oscillation, which we will call KL BBHs in this letter.

Several mechanisms for generating BBH inspirals through KL oscillations have been proposed in the literature. A significant fraction of BBH mergers in LIGO could potentially arise from the KL channel. Therefore it would be interesting to search for GW signals of KL BBHs directly in LISA. A detection of such source will not only reveal the formation channel of the BBHs, but also provide a direct observation of KL mechanism, which was almost never directly observed in astrophysical systems \footnote{The KL oscillation could in principle be searched for in triple systems with a pulsar. However only two such systems are known, one where the second object is too far and the other where it is coplanar, but KL might be observed in the future \cite{Thorsett:1999nc,Ransom:2014xla}. We thank Luc Blanchet for making us aware of these observations.}.

\emph{Kozai-Lidov Oscillations.} Here we review the essential physics of KL BBHs and refer readers to \cite{Randall:2018nud} for a detailed analytical study  and to \cite{2016ARA&A..54..441N} for a review.

We use $m_0$ and $m_1$ to denote the masses of the two binary members, which form the \emph{inner binary}. The inner orbit is in general elliptical, with semi-major axis $a_1(t)$ and eccentricity $e_1(t)$. When a third body with mass $m_2$ is nearby, the inner binary will orbit around $m_2$ along an \emph{outer orbit}, which has semi-major axis $a_2$ and eccentricity $e_2$. We also define $m\equiv m_1+m_2$, $\ep_1\equiv 1-e_1^2$ and $\ep_2\equiv 1-e_2^2$ for later convenience.

At the Newtonian level, the tidal force from the third body $m_2$ could generate long-term variations in the eccentricity and the orientation of the inner orbit. The time scale $t_\text{KL}^{-1}\sim \dot e_1$ of the KL oscillation is
\bge
\label{tKL}
  t_\text{KL}^{-1} \simeq \sqrt{\FR{Gm_2^2a_1^3}{m a_2^6 \ep_1\ep_2^3}},
\ede
where $G$ is Newton's constant. Note that $\ep_1$ can change a lot during a KL cycle and $\ep_1$ in (\ref{tKL}) should be understood as the averaged $\ep_1$ in a KL cycle.  In addition, there are important general relativistic (GR) effects affecting the KL oscillations. First, we have the precession of the orbit at the first order of post-Newtonian correction (1PN), which tends to destroy the KL oscillation. The time scale $t_\text{PN}$ of the precession is,
\bge
  t_\text{PN}^{-1} \simeq \FR{3}{c^2a_1\ep_1}\bigg(\FR{Gm}{a_1}\bigg)^{3/2}.
\ede
Second, the GW radiation from the orbit will reduce the orbital size $a_1$. The time scale $t_\text{GW}$ for this reduction is
\bge
  t_\text{GW}^{-1} \simeq \FR{G^3\mu_1m^2}{c^5a_1^4\ep_1^{5/2}}.
\ede
For the binaries of interest in this paper, at the early stage of the binary evolution, $a_1$ will be large and GR corrections will be small. The KL time scale $t_\text{KL}$ will be the shortest among the above three time scales and thus KL oscillations will initially  dominate the orbital evolution. During this stage,  $e_1(t)$ undergoes oscillations while $a_1(t)$ reduces very slowly. As the binary orbit gets reduced, GR corrections eventually become dominant and quench the KL oscillations. In this final stage the binary feels little tidal perturbation and evolves in isolation. 

\begin{figure*} 
\centering
\includegraphics[width=0.35\textwidth]{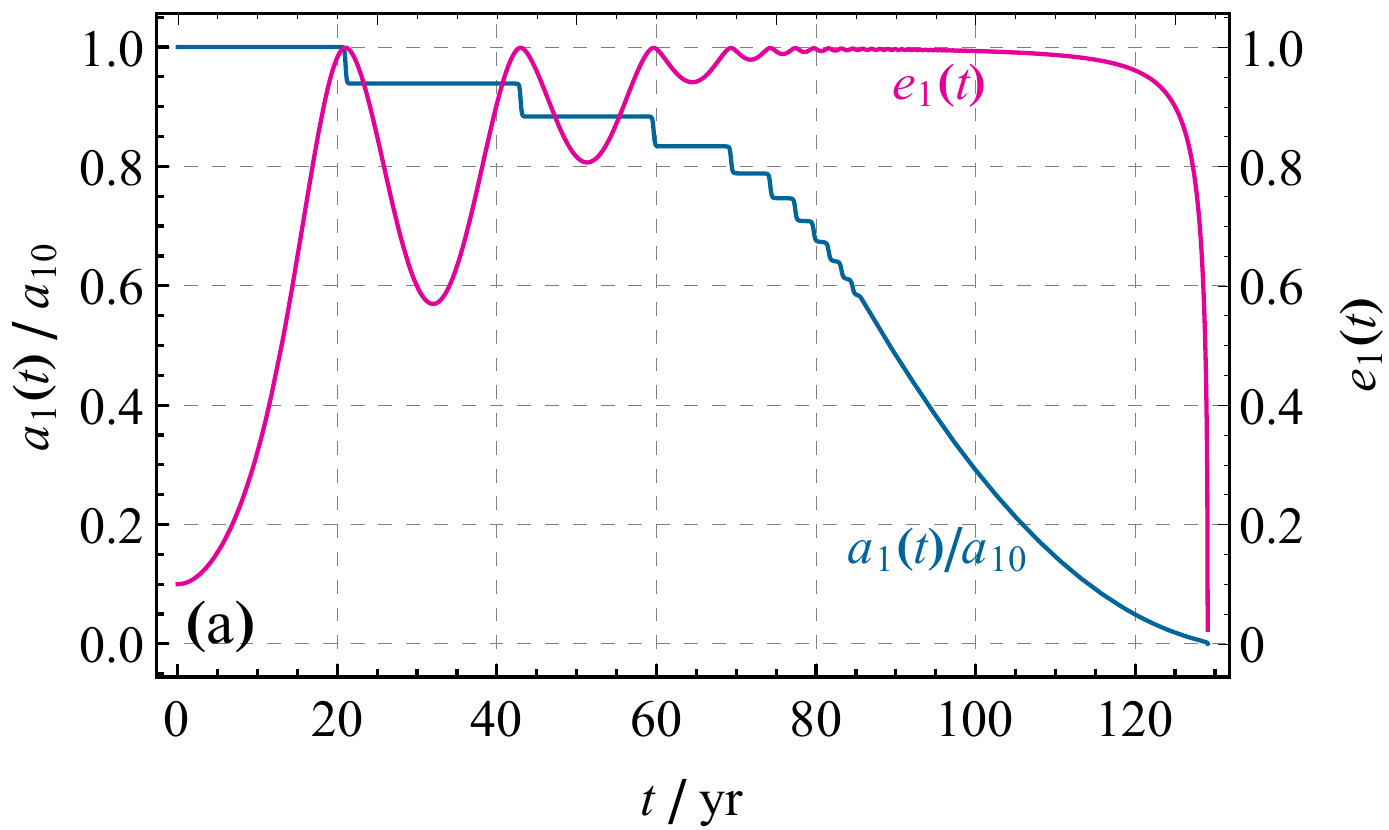}  
\includegraphics[width=0.33\textwidth]{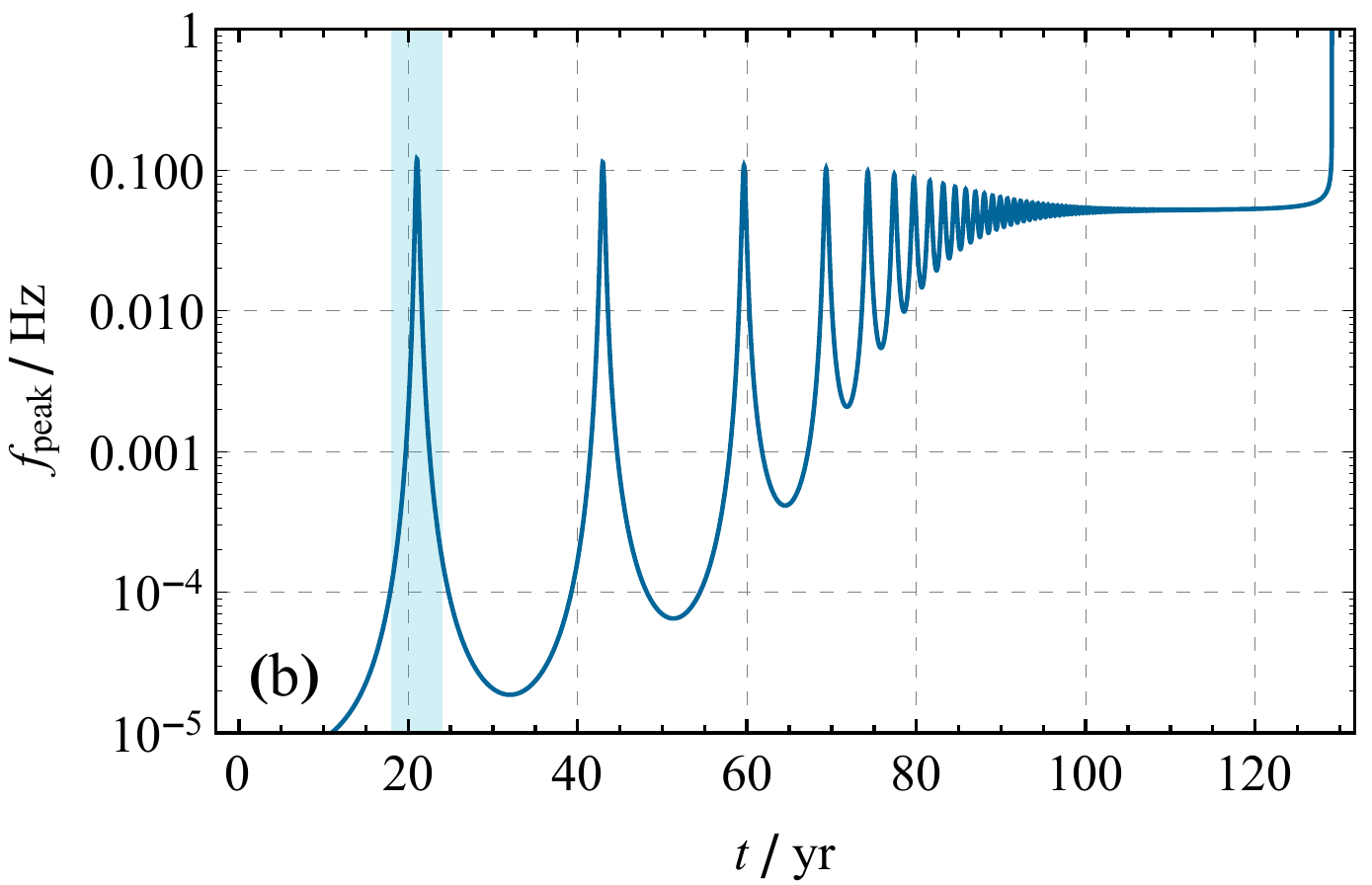}
\includegraphics[width=0.22\textwidth]{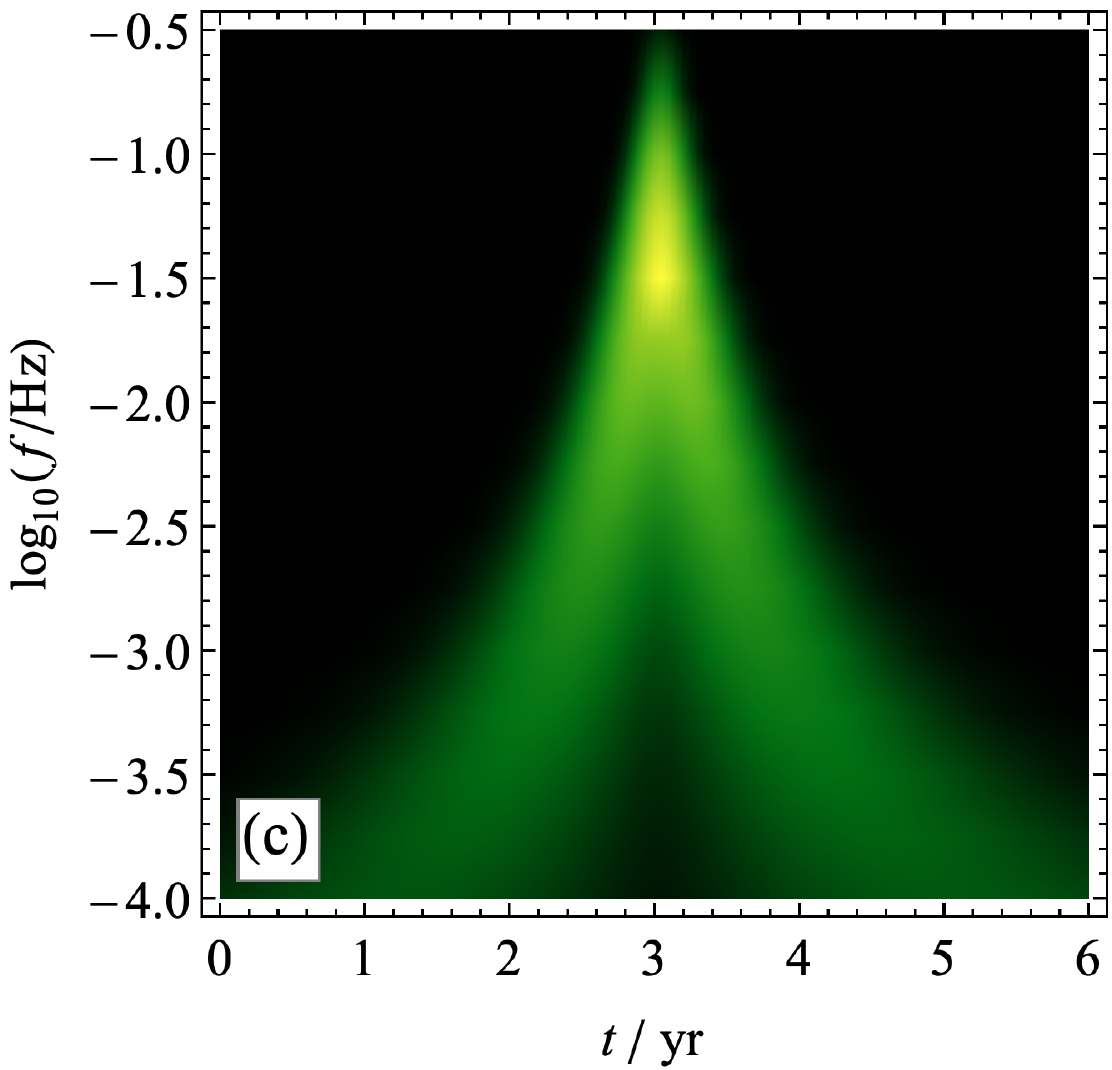} 
\caption{An example of KL BBH in a galactic center. (a) The semi-major axis $a_1(t)$ and the eccentricity $e_1(t)$ as functions of the time. (b) The peak frequency $f_p$ of the GWs emitted by this BBH as a function of time. (c) An illustration of GW spectrum in frequency-time domain, in a 6-year period marked by the shaded strip in (b). In this example, we take $m_0=m_1=30M_\odot$, $m_2=4\times 10^6M_\odot$, $a_{10}=0.2$AU, $a_2$=150AU, $e_{10}=e_2=0.1$ and $I_0=89.9^\circ$. The maximal eccentricity $e_\text{1max}$ reached in this example is $\ep_\text{1min}=1-e_\text{1max}^2=0.0022$ and the merger time is $\tau=129$yr.}
\label{fig_samplesol}
\end{figure*}

In Fig.\;\ref{fig_samplesol} we show a sample of KL BBH in the galactic center. The binary with $m_0=m_1=30M_\odot$ is 150AU away from a supermassive BH with $m_2=4\times 10^6M_\odot$. It starts with $a_{10}=0.2$AU, $e_{10}=0.1$ and $I=89.9^\circ$.

It is possible to capture most of the features of KL BBHs by a simple analytical approximation \cite{Randall:2018nud}.  In short, one can think of the evolution of a KL BBH as a fictitious isolated binary whose GW radiation is quasi-periodically switched off by the KL oscillation. This fictitious binary has identical masses $(m_0,m_1)$ and initial semi-major axis $a_{10}$ to the true system of interest, and has initial eccentricity given by $e_\text{1max}$, which is the maximal eccentricity reached by the KL BBH in its initial KL cycle and is a function of all initial parameters of the binary. This description is valid because GW radiation (and thus the orbital reduction) is most efficient when $e_1$ reaches its maximum. However, the large-$e_1$ configuration is quasi-periodically torqued away by the KL oscillation. Consequently, the KL BBH only spends $\ep_\text{1min}^{1/2}$ ($\ep_\text{1min}\equiv 1-e_\text{1max}^2$) of its time around the large $e_1$. An immediate consequence of this approximation is that the merger time of a KL BBH is given by $\tau=\tau_0\ep_\text{1min}^{-1/2}$, where $\tau_0$ is the merger time of the fictitious binary with initial semi-major axis $a_{10}$ and eccentricity $e_\text{1max}$,
\bge
\label{tau0}
  \tau=\tau_0\ep_\text{1min}^{-1/2},~~~~\tau_0=\FR{5}{256}\FR{c^5a_{10}^4}{G^3m_0m_1m}\ep_\text{1min}^{7/2}.
\ede
The value of $\ep_\text{1min}$ can be determined from the initial parameters of KL BBHs by the energy and angular momentum conservation. An analytical expression of $\ep_\text{1min}$ can be found in \cite{Randall:2018nud}.

\emph{Observability.} In the context of GW detection, previous literature focused almost exclusively on the late stage of GW domination, where $a_1$ is small enough  that the GW frequency falls into the observation band of GW detectors like LIGO or LISA. In this letter we show that in special cases the early stage of evolution when KL dominates the dynamics could also be observed in LISA. This observation is possible if: 1) the GW frequency of such KL BBHs fall in the LISA band; 2) the time scale of eccentricity variation is within the observation time of the LISA mission; 3) there are enough  events with significant signal-to-noise ratio (SNR). Now we consider the three factors in turn.

First, KL BBHs that meet the above criteria typically possess large eccentricity, and therefore emit GWs over a wide range of harmonics. The frequencies of the harmonics $f_n=n\omega/(2\pi)~(n\geq 1)$ are  integer multiples of the orbital frequency $\omega=\sqrt{Gm/a_1^3}$. The GW amplitude $h_n$ of the $n$'th harmonic is related to the radiation power $P_n$ in the $n$'th harmonic by
\bge
\label{hn}
  \la \dot h_n^2\ra = (2\pi f_n)^2\la h_n^2\ra= \FR{4G}{c^3r^2}P_n(a,e),
\ede
The expression of $P_n(a,e)$ for each $n$ is complicated and can be found in \cite{Randall:2017jop}. Here it suffices to note that the spectrum $P_n$ is peaked at the GW frequency $f_p$,
\bge
  \label{fp}
  f_p=\FR{\sqrt{Gm}(1+e_1)^{1.1954}}{\pi (a_1\ep_1)^{3/2}},
\ede
which we shall call the peak frequency. The typical KL BBHs' semi-major axes are much greater than that of the circular BBHs emitting in the LISA band. Therefore its orbital frequency $\omega_0=\sqrt{Gm/a_1^3}$ is much lower than  the LISA band. However, $\ep_1$ can be very small for KL BBHs, and (\ref{fp}) tells us that GWs from a KL BBH would be peaked in the LISA band as long as the combination $a_1\ep_1$ is comparable to the corresponding orbital separation of a circular binary emitting in LISA band. Incidentally, $a_1\ep_1$ is twice of the periapsis distance for large $e_1$.

In Fig.\;\ref{fig_samplesol}(b), we show the peak frequency of the sample KL BBH as a function of time, and in Fig.\ref{fig_samplesol}(c), we show the power of GW radiation over a 6-year period in the frequency-time domain. Due to the change of $e_1$ in time, the emitted GW frequency sweeps over the entire LISA band. It is clear that the signal is vastly different from the chirping of an isolated binary, and the oscillatory behavior in frequency is a direct manifestation of the KL oscillation.  In the next section, we shall show that KL BBHs in dynamical channels will emit GWs typically peaked in the LISA band. Intuitively this is because too high frequency means too fast merging without KL cycles and too low frequency means too weak or no KL oscillations.

Second, the KL oscillations can be seen in LISA only when the eccentricity $e_1(t)$ has significant change during the observation time $T_O\sim\order{1\sim 10}$yr. Note that this does not require that the period of KL oscillation is shorter than $T_O$. In most cases it is unlikely to observe the whole KL cycle. Instead, since small $a_1$ can quench KL oscillations, typical KL BBHs have large $a_1$ and therefore it can be seen in LISA only when $e_1$ is large, too. Around the maximum $e_\text{1max}$, the value of $\ep_\text{1min}$ changes by a factor of 2 during a time $t_{\Delta\ep}=\ep_\text{1min}^{1/2}t_\text{KL}$. This corresponds to the change of $f_p$ by a factor of $2^{-3/2}\sim0.35$ and is thus very significant. 

 From (\ref{tKL}) we see that the time scale $t_{\Delta\ep}$ for this change is,
\begin{align}
  t_{\Delta\ep}=&~4.1\text{yr}\bigg(\frac{\ep_\text{1min}}{0.01}\bigg)^{1/2}\bigg(\frac{m}{60M_\odot}\bigg)^{1/2}\bigg(\frac{30M_\odot}{m_2}\bigg)\n\\
  &~\times\bigg(\frac{\text{AU}}{a_1}\bigg)^{3/2}\bigg(\frac{a_2}{10\text{AU}}\bigg)^3.
\end{align}
This could well be within the LISA observation time. Here we are presenting the result for three $30M_\odot$ objects, a typical case in isolated triple system. One can scale $m_2$ and $a_2$ to find the corresponding time scale when $m_2$ is a supermassive BH in galactic centers. In the galactic-center example in Fig.\;\ref{fig_samplesol}, the corresponding time scale is 0.94yr, which is indeed short enough to be seen in LISA as shown in Fig.\;\ref{fig_samplesol}(c). In the following section, we shall show that an $\order{1}$ fraction of KL BBHs do present significant change in eccentricity when its GWs fall into the LISA band. We note in passing that, even away from the maximum $e_\text{1max}$, the eccentricity variation can still be seen due to the additional phase drift in GW signals \cite{Randall:2018lnh}. 

Third, KL BBHs can be seen in LISA only when there are enough  such sources emitting in the LISA band and with large enough signal-to-noise ratio (SNR). 

The number of KL BBHs per volume, or per Milky-Way-equivalent galaxy (MWEG), can be inferred from the corresponding merger rate $\mathcal{R}=\di n/\di t$ where $n$ is the number density of merging binaries. Then, the differentiatial number density $n$ per frequency interval can be found as
\bge
\label{dndf}
  \FR{\di n}{\di f_p}=\mathcal{R}\dot f_p^{-1}.
\ede
The time dependence of $f_p$ can be found from (\ref{fp}) as $\dot f_p=(\pd f_p/\pd a_1)\dot a_1+(\pd f_p/\pd e_1)\dot e_1$. Here we can use the picture of isolated fictitious binaries, and thus $(\dot a_1,\dot e_1)$ can be found using Peters' equation \cite{Peters:1964zz}. While the full expression for $\dot f_p$ is complicated, the highlight is that the integrated number density $n=\int\di f_p\,\mathcal{R}\dot f_p^{-1}$ is proportional to the merger time $\tau_0$ of binaries in (\ref{tau0}). For fixed $f_p$ and small $\ep_1\ll 1$, we see that $a_1\sim \ep_1^{-1}$ according to (\ref{fp}) and therefore $n\propto \tau_0\propto a_1^4\ep_1^{7/2}\propto \ep_1^{-1/2}$. That is, \emph{the number density of BBHs is increased by a factor of $\ep_1^{-1/2}$ for fixed merger rate.} See also \cite{Fang:2019dnh}. A technical point to be noted here is that $\ep_1$ actually changes with time and thus with $f_p$ for a given BBH. Therefore, to get the number density in (\ref{dndf}) more precisely, we should rewrite $e_1$ as a function of $f_p$, by using (\ref{fp}) and Peters' equations.

\begin{figure*}
\centering
\vcenteredhbox{\includegraphics[height=0.26\textwidth]{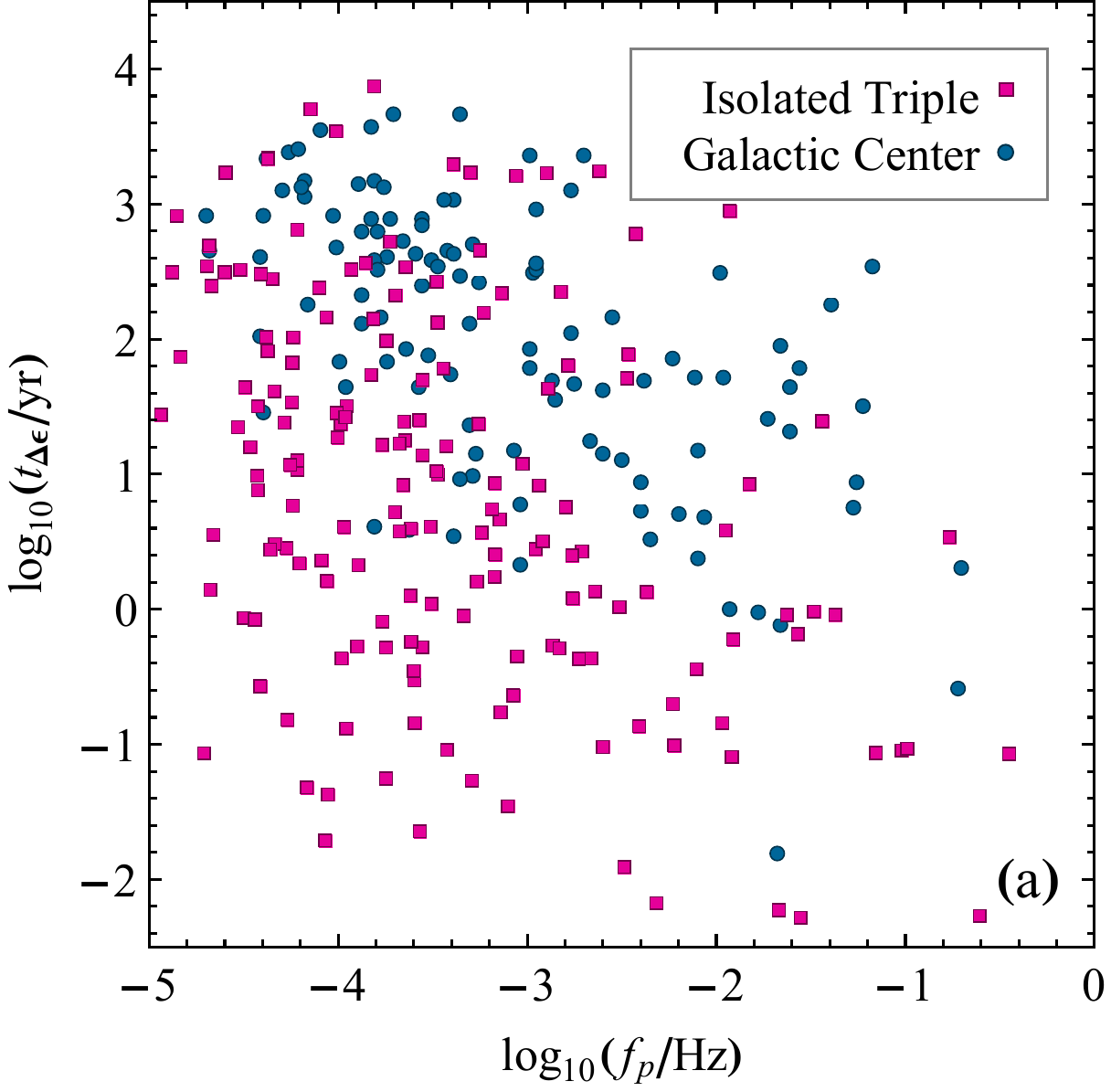}~~~}
\vcenteredhbox{\includegraphics[height=0.265\textwidth]{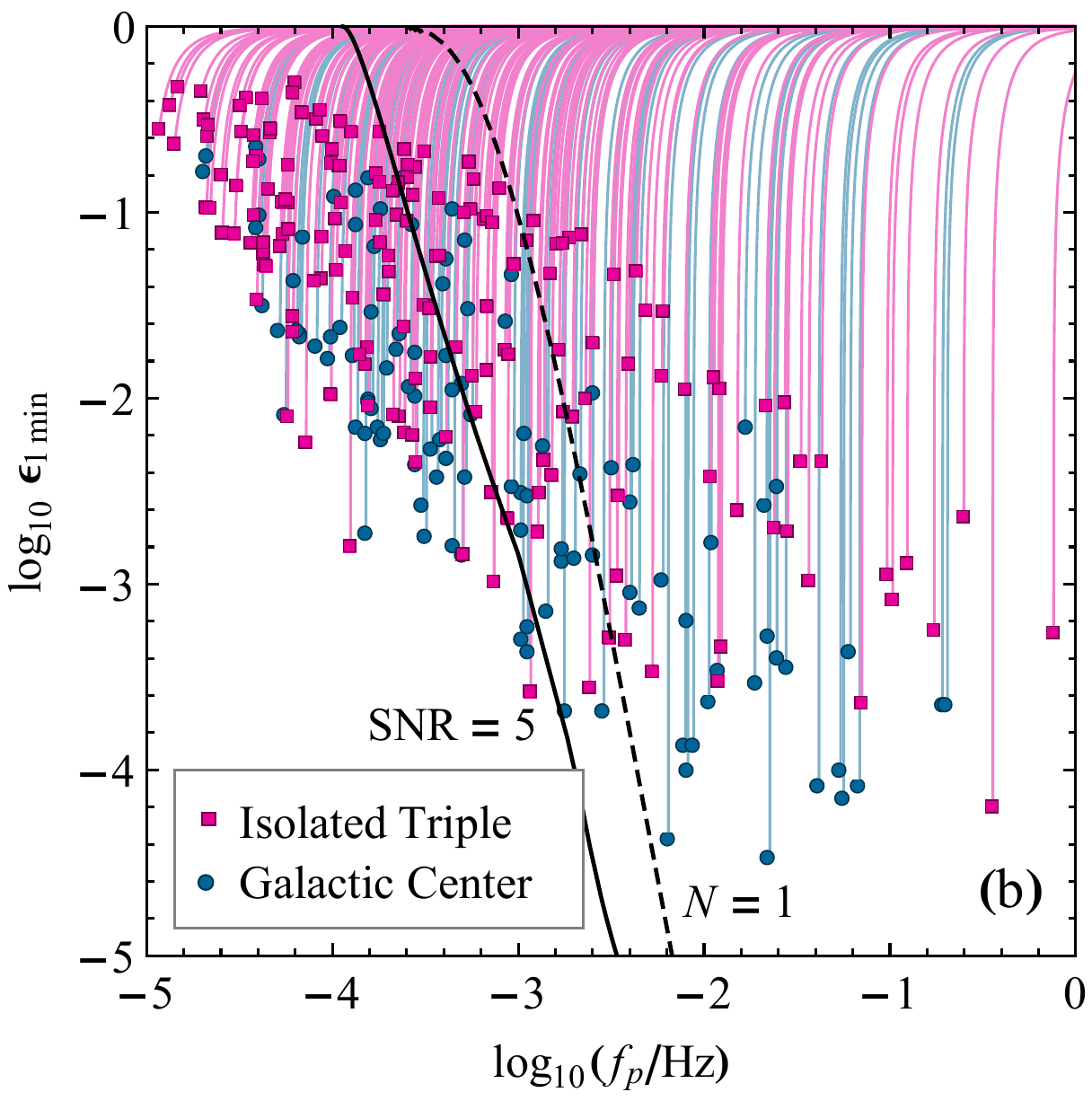}~~~}\vcenteredhbox{\includegraphics[height=0.275\textwidth]{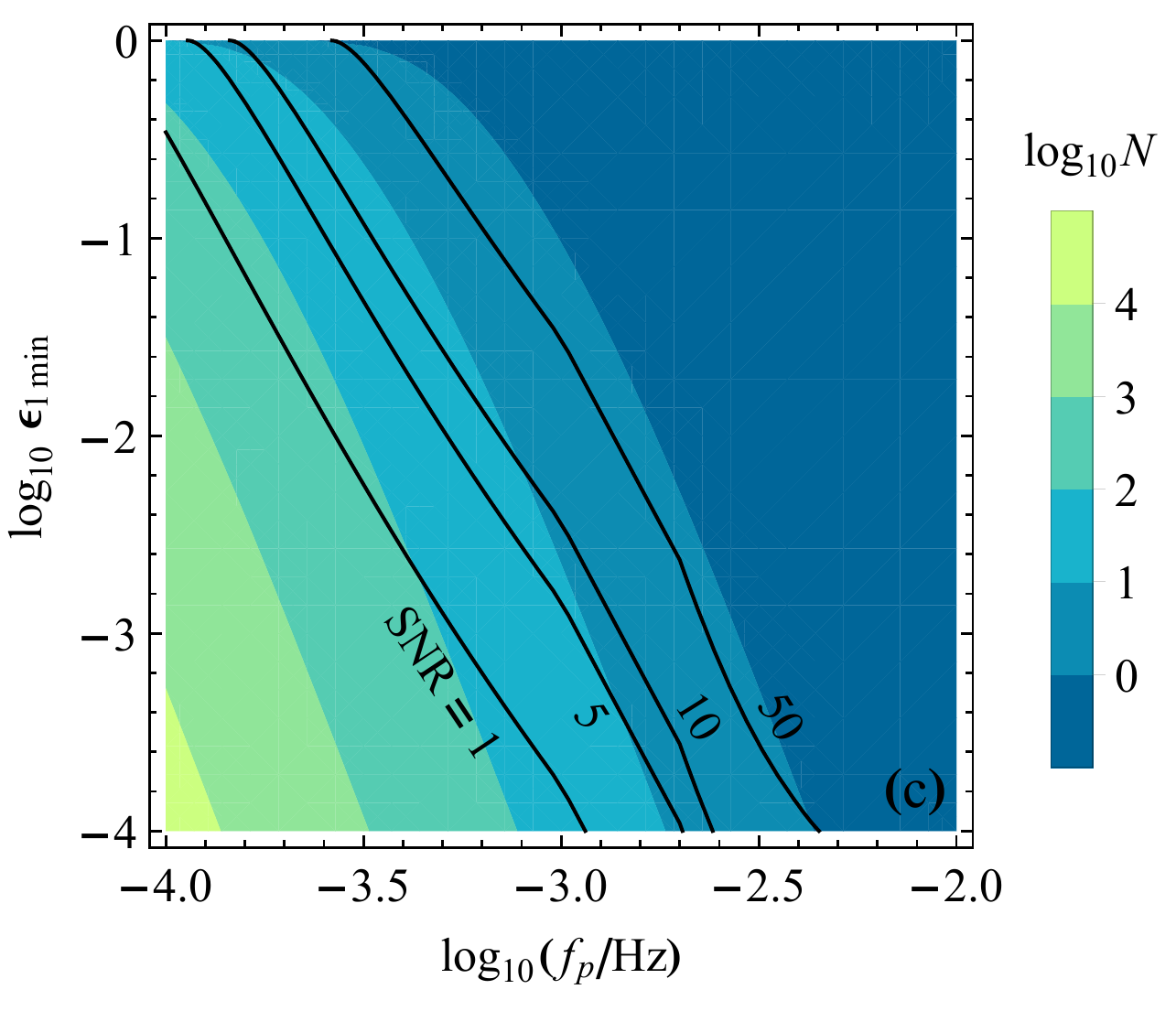}}
\caption{(a) The distribution of $f_p$ and $t_{\Delta\ep}$ during the first KL oscillation, for the sampled KL BBHs in isolated triples (purple) and galactic centers (blue). (b) The distribution of $\ep_\text{1min}$ and $f_p$ during the first KL oscillation for the same samples of KL BBHs in the two channels. The curve attached to each dot describes the subsequent evolution. (c) The number of KL BBHs $N$ (shown in color shadings) at current moment in the Milky Way, assuming that all of them start with the indicated $(f_p,\ep_\text{1min})$, and that KL BBHs contribute 50\% of all BBH mergers. The black curves show the 5-year SNR in LISA of a KL BBHs with $(f_p,\ep_\text{1min})$ at a distance of 10kpc. In (c) we take $m_0=m_1=30M_\odot$. Two contours of SNR$=5$ and $N=1$ from (c) are also shown in (b) for comparison.}
\label{fig_dist}
\end{figure*}

Now we consider the SNR of KL BBHs. The GW signals of KL binaries are more irregular and depend on more parameters  than  isolated binaries. (At least one more parameter $\dot e_1$ is needed.) In addition, large eccentricities introduce more harmonics in GWs, making it likely that a match filtering strategy will be indispensable to dig the signals out of the noise background. All these make the observation of KL binaries more challenging, but also rewarding given the rich physics we can learn from it. Therefore, it would be highly desirable to consider the template making of KL binaries in  future studies. In this work, we assume that the appropriate templates are available and estimate SNR assuming a template search.

To estimate the SNR, we adopt a time-domain approximation, using the noise-weighted waveform $\wh h(t)$ \cite{Barack:2003fp},
\bge
\label{hhat}
  \text{SNR}^2\simeq 4\int_{0}^{T_O}\di t\,\wh h^2(t),~~~\wh h (t)\equiv \sum_{n}\FR{h_n(t)}{S_N^{1/2}(f_n(t))},
\ede
where we have included two independent interferometers of LISA, and $S_N(f)$ is the noise-strain of the GW telescope, for which we take the sky-averaged noise strain with N2A5 configuration \cite{Klein:2015hvg}.  The summation is over the $n$'th harmonics, and the integration is over the whole observation time $T_O$. Using (\ref{hn}), we have,
\bge
  \text{SNR}^2\simeq \FR{G}{\pi^2c^3r^2}\sum_{n}\FR{4}{f_n^2S_N(f_n)} \int_{0}^{T_O}\di t\,P_n(a_1,e_1).
\ede
For large $e_1$ (namely small $\ep_1=1-e_1^2$), it can be shown that the SNR scales with $\ep_1$ as $\ep_1^{3/4}$ for binaries with fixed $f_p$, and the following formula turns out to be a good approximation allowing us to calculate SNR without doing explicit summation over harmonics \cite{RatePaper}.
\bge
\label{snre}
   \text{SNR}(f_p,e_1)\simeq \text{SNR}(f_p,e_1=0)\cdot(1-e_1)^{3/4}. 
\ede
This scaling shows that the SNR decreases for eccentric BBHs compared with a circular BBH emitting at the same frequency. 

Due to the suppression of SNR and the enhancement in the number density, which is very significant for the highly eccentric KL BBHs,  we see that highly eccentric BBHs are likely to be found only at small distances. Therefore it is useful to estimate the number of galactic BBHs with large eccentricity. Given the merger rate $\mathcal{R}=53.2^{+58.5}_{-28.8}\text{Gpc}^{-3}\text{yr}^{-1}$ inferred from LIGO, we can find a rate of merger in our galaxy to be $\mathcal{R}_\text{MW}\simeq 4.58 \text{Myr}^{-1}$, using the conversion $1\text{MWEG}/86\text{Mpc}^3$ \cite{Abbott:2016nhf}.
 
In Fig.\;\ref{fig_dist}(c), we plot the expected number of Milky-Way KL BBHs currently radiating in the LISA window, assuming that KL BBHs contribute half of $\mathcal{R}_\text{MW}$, and that all these KL BBHs were created with initial GW peak frequency $f_p$ and maximal eccentricity $\ep_\text{1min}$. In the same plot we overlay the LISA SNR of a KL BBH with $m_0=m_1=30M_\odot$ at 10kpc, and with GW frequency $f_p$ and eccentricity $\ep_\text{1min}$. This plot shows  that lower $f_p$ and smaller $\ep_\text{1min}$ increase the number of KL BBHs in the galaxy but reduce the SNR in LISA. But there is a region (middle in the plot) with SNR$>$5 and total number of galactic KL BBHs larger than 1 or even 10. As will be detailed in the next section, KL BBHs in several astrophysical channels are expected to populate this region (cf. Fig.\;\ref{fig_dist}b). This shows that LISA has a chance to observe KL BBHs in Milky Way. The scaling of SNR with $\ep_1$ and the estimated Milky-Way BBHs agree well with \cite{Fang:2019dnh}.

\emph{Astrophysical Sources of KL BBHs.} KL BBHs can appear in several proposed dynamical formation channels of compact BBHs. For field triples, the estimated rate depends on the modeling of the triple evolution as well as the distribution of all initial parameters. Ref. \cite{Silsbee:2016djf} reports an estimated rate up to $\mathcal{R}=6\text{Gpc}^{-3}\text{yr}^{-1}$ with a possible reduction from nonzero natal kicks. Ref. \cite{Antonini:2017ash} gives a rate up to 2.5$\text{Gpc}^{-3}\text{yr}^{-1}$ in this channel.

For binaries in SMBH-carrying galactic centers, \cite{Antonini:2012ad} estimated the rate to be 0.048MWEG$^{-1}$Myr$^{-1}$ which is converted to $0.56$Gpc$^{-3}$yr$^{-1}$. The rate could be reduced if there are not enough BBHs, or if the replenished BBHs are very soft that most of them get tidally disrupted at the galactic center.  

The rate of BBH mergers from globular clusters in the local universe have been estimated to be 14Gpc$^{-3}$yr$^{-1}$ \cite{Rodriguez:2018rmd}, and \cite{Rodriguez:2018pss} showed that $\sim 28\%$ of these mergers are from dynamical encounters with a third body. KL oscillations could happen in these BBHs, too.

For the purpose of this work, we estimate the fraction of observable KL BBHs in two channels of isolated triples and galactic centers, which can be treated perturbatively. By observable we mean that KL BBHs should radiate GWs in the LISA window and that their eccentricity variation should be significant enough during the LISA mission. This analysis can also be extended to KL BBHs in other dynamical channels such as those that arise in globular clusters, which we leave for future studies. 

For isolated triples, we sample the initial parameters mostly following \cite{Silsbee:2016djf}. We choose each of $(m_0,m_1,m_2)$ from a log-flat distribution in $[5M_\odot,150M_\odot]$, and choose each of $(a_{10},a_2)$ from a log-flat distribution within $(0.1\text{AU},6000\text{AU})$. We also choose flat distributions of $(e_{10},e_2)$ within [0,1] and of $\cos I_0$ within $[-1,1]$. For a given set of initial parameters, we limit ourselves to binaries merging within the age of galaxies, namely $\tau<10^{10}$ with $\tau$ calculated from (\ref{tau0}). Then, in all merged binaries, we select out all KL BBHs, satisfying the two conditions. First, the KL oscillation should reduce $\ep_1$ from its initial value $\ep_{10}$ by at least a factor of $2$ for it to be significant $\ep_\text{1min}\leq \ep_{10}/2$. Second, the merger time $\tau$ is longer than the KL time scale $t_{KL}$ so that the BBH does not merge within one KL cycle. This ensures that there are at least several KL oscillations before the coalescence. 

For BBHs in galactic centers, we choose $(m_0,m_1)$ in a log-flat distribution in $[5M_\odot,150M_\odot]$ and $a_1$ in a log-flat distribution in $[0.1\text{AU},100\text{AU}]$. We fix $m_2=4\times 10^6M_\odot$, and  $a_2$ follows a mass-segregated distribution from 1AU to 0.1pc. Both of $(e_{10},e_2)$ follows the thermal distribution, namely a flat distribution in $e_{1,2}^2$, and $\cos I_0$ is flat in $[-1,1]$. In this channel we remove tidally disrupted and evaporated BBHs and then pick up the KL BBHs.  

We generate 500 mergers each time from isolated triples and from the galactic center, respectively, and repeat 20 times. We find that, of all merging binaries, $(39\pm2)\%$ of isolated-triple channel, and $(22\pm 2)\%$ of galactic-center channel, exhibit significant KL oscillations. We have taken the double-average approximation at the quadrupole level. Since the non-secular effects turn out to be important in the isolated triples, the fraction of KL BBHs here is likely an underestimate.

In Fig.\;\ref{fig_dist} we show the resulting unnormalized distribution of KL BBHs in the planes of $(f_p,t_{\Delta\ep})$ and of $(f_p,\ep_\text{1min})$. The dots represent the value during the first KL cycle. In Fig.\;\ref{fig_dist}(a), we see an large fraction of KL BBHs are in the observable window with $0.1\text{mHz}<f_p<0.1$Hz and $t_{\Delta\ep}\lesssim \order{10\text{yr}}$. Fig.\;\ref{fig_dist}(b) shows that the KL BBHs typically have large eccentricity, which is needed to speed up the merger for initially wide binaries.  A comparison between Fig.\;\ref{fig_dist}(b) and Fig.\;\ref{fig_dist}(c) shows that many of these KL BBHs can have significant SNR in LISA if they are in the Milky Way, and, there can be several to tens of such KL BBHs in the Milky Way, if the KL channel contributes $50\%$ of merger events in LIGO. Even if we reduce the contribution of KL BBHs to 5\%, there are still up to several such KL BBHs currently radiating in the LISA band and their eccentricity variation in LISA could be observable.

\emph{Discussion.} In this letter we showed that the KL oscillation of a BBH could be observed through GWs in LISA. If KL BBHs contribute a significant fraction of BBH mergers in the universe, there could be several to tens of KL BBHs from our own galaxy radiating in the LISA band. A discovery of KL BBHs would not only reveal the formation channel of the BBHs, but also be a direct observation of KL oscillations. 

The observation of KL BBHs relies crucially on the templates or other methods like stacking for observing highly eccentric binaries and also with time variations in the eccentricity. Therefore it would be desirable to consider the templates or stacking method for these signals in the future.

In addition, LISA can resolve up to thousands of galactic binaries \cite{Audley:2017drz}. Many of them could be in triple systems and one can also look for KL oscillations in these systems. We leave this possibility for future studies.

\begin{acknowledgements} 
LR was supported by an NSF grant PHY-1620806, a Kavli Foundation grant ``Kavli Dream Team,'' the Simons Fellows Program, 
the Guggenheim Foundation, and an IHES CARMIN fellowship. LR thanks the Institute Henri Poincaré for their hospitality when this work was initiated.
\end{acknowledgements}


%

\end{document}